# DRetNet: A Novel Deep Learning Framework for Diabetic Retinopathy Diagnosis

Idowu Paul Okuwobi*, *Senior Member, IEEE*, Jingyuan Liu*, Jifeng Wan, and Jiaojiao Jiang

***Abstract*—Diabetic retinopathy (DR) is a leading cause of blindness worldwide, necessitating early detection to prevent vision loss. Current automated DR detection systems often struggle with poor-quality images, lack interpretability, and insufficient integration of domain-specific knowledge. To address these challenges, we introduce a novel framework that integrates three innovative contributions: (1) Adaptive Retinal Image Enhancement Using Physics-Informed Neural Networks (PINNs): this technique dynamically enhances retinal images by incorporating physical constraints, improving the visibility of critical features such as microaneurysms, hemorrhages, and exudates; (2) Hybrid Feature Fusion Network (HFFN): by combining deep learning embeddings with handcrafted features, HFFN leverages both learned representations and domain-specific knowledge to enhance generalization and accuracy; (3) Multi-Stage Classifier with Uncertainty Quantification: this method breaks down the classification process into logical stages, providing interpretable predictions and confidence scores, thereby improving clinical trust.**

**Comprehensive evaluations demonstrate significant improvements in accuracy, robustness, and interpretability. The proposed framework achieves an accuracy of 92.7%, a precision of 92.5%, a recall of 92.6%, an F1-score of 92.5%, an AUC of 97.8%, a mAP of 0.96, and an MCC of 0.85. Ophthalmologists rated the framework's predictions as highly clinically relevant (4.8/5), highlighting its alignment with real-world diagnostic needs. Qualitative analyses, including Grad-CAM visualizations and uncertainty heatmaps, further enhance the interpretability and trustworthiness of the system. The framework demonstrates robust performance across diverse conditions, including low-quality images, noisy data, and unseen datasets. These features make the proposed framework a promising tool for clinical adoption, enabling more accurate and reliable DR detection in resource-limited settings.**

## I. Introduction

DIABETIC retinopathy (DR) is a leading cause of blindness among working-age adults globally, affecting millions of individuals with diabetes [1]. It is characterized by damage to the blood vessels in the retina, leading to vision impairment and potential blindness if left untreated [2]. Fig. 1 shows different severity of DR associated with different characteristics. Early detection and timely intervention are critical for preventing irreversible vision loss [3]. Despite significant advancements in medical technology, DR remains a major public health concern, particularly in resource-limited settings where access to specialized care is limited [4]. Traditional DR screening involves manual examination by ophthalmologists, which is time-consuming, labor-intensive, and prone to human error [5]. As a result, automated DR detection systems have gained considerable attention in recent years. These systems leverage computer vision and machine learning techniques to analyze retinal images and classify the severity of DR [6]. However, existing automated systems face several challenges, including: (1) Poor-Quality Images: retinal images often suffer from uneven illumination, noise, and artifacts, which can significantly degrade the performance of automated systems [7]; (2) Limited Interpretability: many deep learning models act as "black boxes," making it difficult for clinicians to understand and trust the predictions [8]; (3) Over-Reliance on Deep Learning: current systems often neglect domain-specific knowledge, such as blood vessel patterns, texture features, and optic disc localization, which are critical for accurate diagnosis [9].

Recent advancements in deep learning have significantly improved the accuracy of DR detection systems. Gulshan et al. [10] demonstrated high accuracy using Inception-v3 for DR grading. Similarly, Rajpurkar et al. [11] introduced CheXNet, a deep learning model for chest X-ray diagnosis, which inspired the development of similar architectures for retinal image analysis [12]. Other researchers [13], [14], [15],

" This work was supported by the National Natural Science Foundation of China (62250410370), National Science and Technology Funding for Foreign Scholar Research Fund Project (WGXZ2023071L), National Science and Technology Funding for Foreign Youth Talent Program (QN2022033002L), Guangxi Natural Science Foundation for Youth Science and Technology (2021GXNSFBA220075)." *(Corresponding author: I.P. Okuwobi and J. Liu).*

I.P. Okuwobi is with School of Life & Environmental Sciences, Guilin University of Electronic Technology, Guilin, Guangxi 541004, China. He is also with Nantong Hamadun Medical Technology Co., Ltd, Nantong, Jiangsu 226400, China (Email: paulokuwobi@ieee.org).

J. Liu is with Nantong Hamadun Medical Technology Co., Ltd, Nantong, Jiangsu 226400, China (Email: 3656124302i@qq.com).

J. Wan and J. Jiang are with Department of Ophthalmology, The Affiliated Hospital of Guilin Medical University, Guilin, Guangxi Province 541001, China.

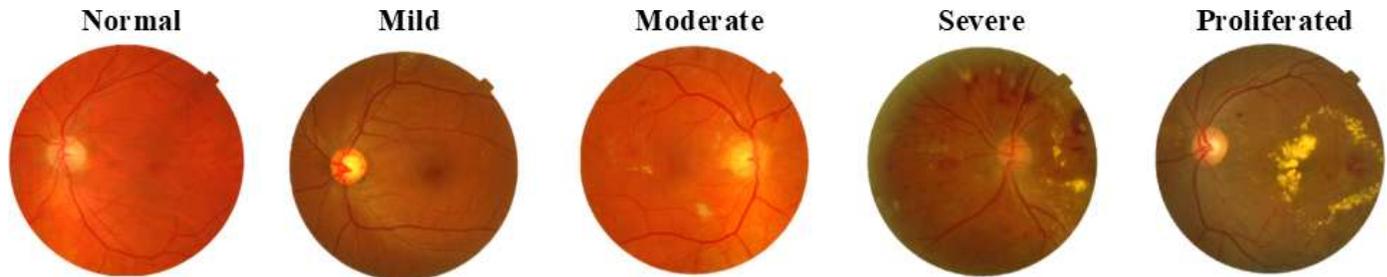

**Fig. 1** Images of DR with different severity levels. Each of these severity level possess different characteristics and features.

**Table 1** Summary of Dataset usage. These datasets are widely recognized in the field of medical imaging and DR research, offering diverse retinal images with varying quality levels, lighting conditions, and clinical annotations.

| DATASET | PURPOSE | NO. OF IMAGES | IMAGE RES. RANGE | CLASS DISTR. |
|---|---|---|---|---|
| Messidor-2 | Validation & Testing | 1,748 | 1440*960*2304*1536 | 0: 60%, 1: 20%, 2: 10%, 3: 5%, 4: 5% |
| Kaggle DR Dataset | Training & Tuning | 35,126 | 1024*1024*4288*2848 | 0: 73%, 1: 18%, 2: 6%, 3: 2%, 4: 1% |
| iDRiD | Cross-Dataset Testing | 516 | 1024*1024*4288*2848 | 0: 50%, 1: 25%, 2: 15%, 3: 7%, 4: 3% |

[16], [17], [18], [19], [20], [21], [22], [23] have also proposed several other techniques. However, these models often struggle with poor-quality images and lack interpretability [24]. To address these issues, researchers have explored various techniques such as: (1) Image Enhancement: techniques such as histogram equalization, contrast stretching, and denoising filters have been used to improve image quality [25]. However, these methods are often insufficient for complex cases, such as those with uneven illumination or severe noise; (2) Explainable AI (XAI): methods like Grad-CAM [26] and LIME [27] have been employed to visualize and explain model predictions, enhancing trust in automated systems [28]; (3) Domain-Specific Features: integrating handcrafted features, such as blood vessel maps and texture descriptors, has been shown to improve model performance [29]. However, these features are often overlooked in deep learning-based approaches [30]. Despite these advancements, there remains a gap in integrating physics-informed preprocessing, hybrid feature fusion, and uncertainty quantification into a unified framework.

Our proposed framework addresses these gaps by combining these techniques to achieve superior accuracy, robustness, and interpretability. The proposed framework aims to overcome the limitations of existing systems by: (a) Improving Image Quality: enhancing the visibility of clinically relevant features, such as microaneurysms, hemorrhages, and exudates, which are critical for accurate diagnosis; (b) Enhancing Interpretability: providing clinicians with actionable insights through Grad-CAM (Gradient-weighted Class Activation Mapping) visualizations and uncertainty heatmaps; (c) Ensuring Robustness: demonstrating strong performance across diverse conditions, including low-quality images, noisy data, and unseen datasets. The main contributions of this work are:

1. Adaptive Retinal Image Enhancement Using Physics-Informed Neural Networks (PINNs): We propose a novel method for enhancing retinal images using physics-informed neural networks, ensuring that the enhanced images adhere to optical principles.
2. Hybrid Feature Fusion Network (HFFN): We introduce a hybrid feature fusion network that combines deep learning embeddings with handcrafted features, improving generalization and accuracy.
3. Multi-Stage Classifier with Uncertainty Quantification: We develop a multi-stage classifier that provides interpretable predictions with confidence scores, enhancing clinical trust and decision-making.

## II. DATASETS

A total of 37,390 images were used in this study. We utilized three publicly available datasets: Messidor-2, the Kaggle Diabetic Retinopathy Detection Dataset, and the IDRiD (Indian Diabetic Retinopathy Image Dataset). Table 1 provide a detail description of the datasets, including its composition and usage in our experiments. Each image is labeled with one of five DR severity grades (0-4): (a) 0: No DR (Healthy), (b) 1: Mild Non-Proliferative DR (NPDR), (c) 2: Moderate NPDR, (d) 3: Severe NPDR, (e) 4: Proliferative DR.

## III. METHOD

Fig. 2 shows the architecture of the proposed DRetNet. The raw retinal images are first normalized and resized. The images are subsequently enhanced using a novel adaptive retinal image enhancement network. Deep and handcrafted features were simultaneously extracted from the images. The proposed hybrid features fusion architecture then fuse the features using multi-head attention and a multi-class classification operation based on the proposed uncertainty quantification method to classify the images. The post-processing operation includes; generation of Grad-CAM visualization, and uncertainty map to guide the final prediction confidence scores. We detail the architecture of the proposed DRetNet in the following sections.

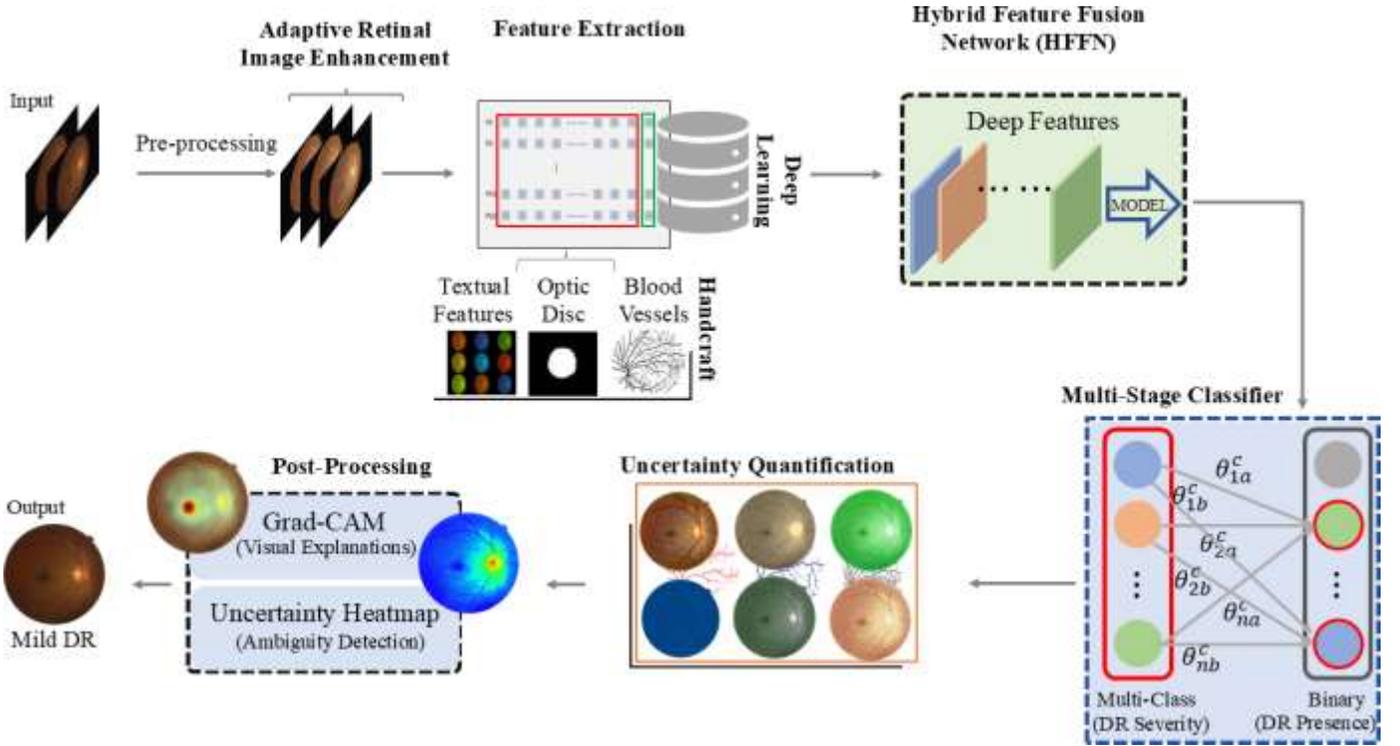

**Fig. 2** The architecture of the proposed DRetNet, a novel framework for DR detection.

### A. Preprocessing

Preprocessing plays a critical role in preparing raw retinal images for analysis by the proposed framework. Poor-quality images with uneven illumination, noise, or artifacts can significantly degrade the performance of DR detection systems. The preprocessing steps aim to normalize, enhance, and standardize the input data to ensure optimal performance of the subsequent components. Given an input image $I$ with pixel values $I_{i,j}$, normalization is performed as follows:

$$I'_{i,j} = \frac{I_{i,j} - min(I)}{max(I) - min(I)} \quad (1)$$

where $I_{i,j}$ is the pixel value at position $(i,j)$, $min(I)$ and $max(I)$ are the minimum and maximum pixel values in the image, respectively, $I'_{i,j}$ is the normalized pixel value. After the normalization operation, we resized the images to a fixed resolution, typically 224*224 pixels using bilinear interpolation.

### B. Adaptive Retinal Image Enhancement

Retinal images often suffer from poor quality due to uneven illumination, noise, or artifacts, which can significantly degrade the performance of DR detection systems. Traditional image enhancement techniques, such as histogram equalization and denoising filters, are often insufficient for complex cases. To address these challenges, we propose an adaptive enhancement method using PINNs. The proposed adaptive enhancement method incorporates physical constraints into the neural network architecture, ensuring that the enhanced images adhere to optical principles and improve the visibility of clinically relevant features. Given a low-quality retinal image $I_{low}$, the goal is to learn a mapping $f_\theta$ that enhances the image to produce a high-quality image $I_{high}$ as:

$$f_\theta(I_{low}) \approx I_{hih} \quad (2)$$

where $\theta$ represents the trainable parameters of the PINN. To ensure that the enhanced images adhere to optical principles, we incorporated physical constraints, particularly the Beer-Lambert Law for light absorption. The Beer-Lambert Law states that the intensity of light decreases exponentially with the distance it travels through a medium:

$$I(d) = I_0 e^{-\mu d} \quad (3)$$

where $I(d)$ is the intensity of light at the distance $d$, $I_0$ is the initial intensity of light, $\mu$ is the absorption coefficient, and $d$ is the path length of light through the medium. In the context of retinal images, we modified the Beer-Lambert Law to model the relationship between the low-quality image $I_{low}$ and the high-quality image $I_{high}$ as:

$$log\left(\frac{I_{high,i}}{I_{low,i}}\right) \approx \mu d_i \quad (4)$$

where $I_{high,i}$ and $I_{low,i}$ are the intensities of the high-quality and low-quality images at the pixel $i$, $d_i$ is the path length of light through the retina at the pixel $i$, and $\mu$ is the absorption coefficient, which is assumed to be constant across the image.

We propose a reconstruction loss to measures the pixel-wise differences between the enhanced image $I_{enhanced}$ and the high-quality image $I_{high}$ as:

$$L_{recon} = \frac{1}{N} \sum_{i=1}^{N} (I_{enhanced,i} - I_{high,i})^2 \quad (5)$$

where $N$ is the total number of pixels in the image, $I_{enhanced}$ and $I_{high,i}$ are the intensities at the pixel $i$ in the enhanced and

2high-quality images, respectively. The physics-informed loss enforces the Beer-Lambert Law by penalizing deviations from the expected relationship between the high-quality and low-quality images as:

$$L_{physics} = \frac{1}{N}\sum_{i=1}^{N}\left(log\left(\frac{I_{high,i}}{I_{low,i}}\right) - \mu d_i\right)^2 \quad (6)$$

where $\mu$ is the absorption coefficient, $d_i$ is the path length of light through the retina at pixel $i$. The total loss function combines the reconstruction loss and the physics-informed loss, weighted by a hyperparameter $\lambda$ as:

$$L_{total} = L_{recon} + \lambda L_{physics} \quad (7)$$

where $\lambda$ controls the trade-off between the reconstruction loss and the physics-informed loss.

## C. Feature Extraction

Feature extraction is a critical step in the proposed framework, providing the necessary inputs for subsequent stages such as hybrid feature fusion and classification. The proposed DRetNet employs two types of feature extraction methods: deep learning-based and handcrafted feature extraction. These two feature sets are later fused using a multi-head attention mechanism to leverage the strengths of both approaches. Deep learning models, especially Convolutional Neural Networks (CNNs), excel at capturing high-level semantic features from images. These features include abstract patterns such as edges, textures, and shapes, which are essential for accurately detecting DR. To leverage these capabilities, we use a pre-trained CNN, specifically ResNet-50, to extract deep learning embeddings from the enhanced retinal images. Each residual block in ResNet-50 is defined by:

$$F(x) = \sigma(W_2 \cdot ReLU(W_1 \cdot x + b_1) + b_2) \quad (8)$$

where $x$ is the output to the block, $W_1, W_2, b_1$, and $b_2$ are the weights and biases of the convolutional layers, and $\sigma$ is the activation function. The output of the block is:

$$y = F(x) + x \quad (9)$$

The residual connection $x$ ensures that the identity mapping is preserved. Given an enhanced retinal image $I_{enhanced}$, the deep learning features $F_{deep}$ are extracted as follows:

1. Pass the image through the pre-trained ResNet-50:
$$F_{deep} = ResNet - 50(I_{enhanced}) \quad (10)$$
2. Extract the activations from the penultimate layer (before the classification layer): $F_{deep} \in \mathbb{R}^{d_{deep}}$
   where $d_{deep}$ is the dimensionality of the feature vector (2048 for ResNet-50).

While deep learning features capture high-level semantic information, they may overlook domain-specific details such as blood vessel patterns, texture, and optic disc localization. To address this, we extract handcrafted features using traditional image processing techniques. These are the handcrafted features extracted:

1. Blood Vessel Maps: Blood vessels are critical indicators of DR severity. We use a vessel segmentation algorithm to extract binary maps of blood vessels:
$$M_{vessels} = SegmentVessel(I_{enhanced}) \quad (11)$$
   where $M_{vessels}$ is a binary mask indicating the presence of blood vessels.
2. Textual Features: Texture features capture local patterns in the image. We compute Haralick texture features using the gray-level co-occurrence matrix (GLCM):
$$T_{texture} = HaralickFeature(I_{enhanced}) \quad (12)$$
   where $T_{texture}$ includes features such as contrast, correlation, energy, and homogeneity.
3. Optic Disc Localization: The optic disc is a key anatomical structure in retinal images. We localize the optic disc using a circular Hough transform:
$$(x_{disc}, y_{disc}, r_{disc}) = HoughTransform(I_{enhanced}) \quad (13)$$
   where $(x_{disc}, y_{disc})$ are the coordinates of the optic disc center, and $r_{disc}$ is its radius.

## D. Hybrid Feature Fusion Network (HFFN)

Traditional deep learning models excel at capturing high-level semantic features but may overlook domain-specific details such as blood vessel patterns, texture, and optic disc localization. Conversely, handcrafted features explicitly encode these domain-specific details but lack the ability to capture complex, abstract patterns. To address this gap, we propose a Hybrid Feature Fusion Network (HFFN) that fuses deep learning embeddings with handcrafted features using a multi-head attention mechanism. This approach dynamically weighs the importance of each feature set based on the input, ensuring optimal utilization of both types of features. Let $F_{deep} \in \mathbb{R}^{d_{deep}}$ and $F_{handcrafted} \in \mathbb{R}^{d_{handcrafted}}$ denote the deep learning and handcrafted features, respectively. The multi-head attention mechanism computes the fused features $F_{fused}$ as follows:

1. Compute query, key, and value matrices:
$$Q = W_Q \cdot [F_{deep}, F_{handcrafted}]$$
$$K = W_K \cdot [F_{deep}, F_{handcrafted}]$$
$$V = W_V \cdot [F_{deep}, F_{handcrafted}] \quad (14)$$
   where $W_Q, W_K$, and $W_V$ are learnable weight matrices.
2. Compute attention scores:
$$Attention(Q, K, V) = SoftMax\left(\frac{Q \cdot K^T}{\sqrt{d_k}}\right) \quad (15)$$
   where $d_k$ is the dimensionality of the key vectors.
3. Combine outputs from multiple heads:
$$F_{fused} = Concat(Head_1, Head_2, \dots Head_3) \cdot W_o \quad (16)$$

The hybrid features fusion network leverages a multi-head attention mechanism to combine deep learning embeddings and handcrafted features. This approach ensures that the model captures both high-level semantic information and domain-specific details, improving its performance in detecting DR.

## E. Multi-Stage Classifier

The multi-stage classifier is a critical component of the proposed framework for DR detection. It consists of two stages: a binary classifier to detect the presence of DR and a multi-class classifier to classify the severity of DR. Additionally, the classifier incorporates uncertainty quantification using Monte Carlo Dropout to provide confidence scores and highlight ambiguous cases. Current DR classifiers often provide binary or multi-class predictions without explaining uncertainty or intermediate reasoning. This lack of interpretability can hinder trust in automated systems.

To address this, we introduce a multi-stage classifier that: (a) detects DR presence (binary classification), and (b) classifies severity levels (multi-class classification). Uncertainty quantification provides clinicians with confidence scores and highlights ambiguous cases for manual review, enhancing trust and decision-making. The binary classifier uses a feedforward neural network with a single output unit and a sigmoid activation function as:

$$y_{binary} = \sigma(W_1 F_{fused} + b_1) \quad (17)$$

where $W_1 \in \mathbb{R}^{1 \times d_{fused}}$ is the weight matrix, $b_1 \in \mathbb{R}$ is the bias term, $\sigma$ is the sigmoid activation functions: $\sigma(z) = \frac{1}{1+e^{-z}}$. The binary cross-entropy loss is used to train the binary classifier as:

$$L_{binary} = -[y \log(y_{binary}) + (1-y) \log(1 - y_{binary})] \quad (18)$$

where $y \in \{0,1\}$ is the true binary label.

The multi-class classifier uses a feedforward neural network with a SoftMax activation function as:

$$y_{multi-class} = SoftMax(W_2 F_{fused} + b_2) \quad (19)$$

where $W_2 \in \mathbb{R}^{5 \times d_{fused}}$ is the weight matrix, $b_2 \in \mathbb{R}^5$ is the biased vector, $SoftMax(z)$ is defined as:

$$SoftMax(z)_i = \frac{e^{z_i}}{\sum_{j=1}^{5} e^{z_j}} \quad (20)$$

where $z = W_2 F_{fused} + b_2$.

The categorical cross-entropy loss is used to train the multi-class classifier as:

$$L_{multi-class} = -\sum_{i=1}^{5} y_i \log(y_{multi-class}) \quad (21)$$

where $y \in \{0,1\}^5$ is the true multi-class label, with exactly one element equal to 1 (indicating the true class).

The multi-stage classifier with uncertainty quantification provides a robust and interpretable solution for diabetic retinopathy detection. By combining binary and multi-class classification with Monte Carlo Dropout, the classifier ensures accurate and trustworthy predictions, making it a valuable tool for clinical practice.

### F. Uncertainty Quantification

Deep learning models are often seen as "black boxes," making it challenging for clinicians to understand and trust their predictions. To address this, we incorporate uncertainty quantification using Monte Carlo Dropout. This approach allows us to estimate the model's confidence in its predictions and identify cases where manual review may be necessary. We utilized two uncertainties: (a) Aleatoric Uncertainty: captures noise inherent in the data (e.g., poor-quality images), (b) Epistemic Uncertainty: captures uncertainty due to the model's parameters (e.g., limited training data). By quantifying uncertainty, the system highlights cases where the model is less confident, allowing clinicians to prioritize these cases for further review.

The goal of uncertainty quantification is to estimate the predictive distribution $p(y|x, \mathcal{D})$, where: $y$ is the predicted label (binary or multi-class), $x$ is the input image, $\mathcal{D}$ is the training dataset. Using Monte Carlo Dropout, the predictive distribution is approximated as:

$$p(y|x, \mathcal{D}) \approx \frac{1}{T} \sum_{t=1}^{T} p(y|x, \theta_t) \quad (22)$$

where $T$ is the number of Monte Carlo samples, and $\theta_t$ represents the model parameters sampled during the $t$-th forward pass with dropout enabled. From the predictive distribution, we compute the mean prediction $\mu_y$ and variance $\sigma_y^2$ as:

$$\mu_y = \mathbb{E}[y] = \frac{1}{T} \sum_{t=1}^{T} y_t \quad (23)$$

$$\sigma_y^2 = Var[y] = \frac{1}{T} \sum_{t=1}^{T} (y_t - \mu_y)^2 \quad (24)$$

The variance $\sigma_y^2$ quantifies the model's uncertainty. Higher variance indicates greater uncertainty in the prediction. For binary classification, the model predicts the probability of DR presence as:

$$p(y=1|x, \mathcal{D}) \approx \frac{1}{T} \sum_{t=1}^{T} \sigma(W_1 F_{fused} + b_1) \quad (25)$$

where $\sigma$ is the sigmoid activation function, $W_1$ and $b_1$ are the weight matrix and bias vector for the binary classifier. The mean prediction and variance are computed as:

$$\mu_{binary} = \frac{1}{T} \sum_{t=1}^{T} y_t \quad (26)$$

$$\sigma_{binary}^2 = \frac{1}{T} \sum_{t=1}^{T} (y_t - \mu_{binary})^2 \quad (27)$$

For multi-class classification, the model predicts the probabilities for each severity level as:

$$p(y=c|x, \mathcal{D}) \approx \frac{1}{T} \sum_{t=1}^{T} SoftMax(W_2 F_{fused} + b_2)_c \quad (28)$$

where $c$ is the class index, and $W_2$ and $b_2$ are the weight matrix and bias vector for the multi-class classifier.

The mean prediction and variance for each class are computed as:

$$\mu_c = \frac{1}{T} \sum_{t=1}^{T} y_{t,c} \quad (29)$$

$$\sigma_c^2 = \frac{1}{T} \sum_{t=1}^{T} (y_c - \mu_c)^2 \quad (30)$$

The proposed uncertainty quantification using Monte Carlo Dropout provides a principled way to estimate predictive uncertainty in the multi-stage classifier. By computing the mean prediction and variance across multiple forward passes, the system offers interpretable confidence scores, improving trust and reliability in clinical settings.

### G. Post-Processing Operation

Post-processing is essential for improving the interpretability and trustworthiness of automated DR detection systems. By visualizing regions contributing to predictions and highlighting areas of uncertainty, clinicians can better understand and validate the model's decisions. This is particularly important in medical applications where interpretability is paramount. Grad-CAM provides visual explanations by highlighting regions in the input image that contribute most to the model's predictions. This helps clinicians identify clinically relevant features such as microaneurysms, hemorrhages, and exudates. We compute the gradients of the final class score $y_c$ with respect to the feature map $A$ of the last convolutional layer as:
$$\frac{\partial y_c}{\partial A} \quad (31)$$

Then, global average pooling is applied to obtain a class activation map (CAM) as:
$$CAM(x,y) = \sum_{k=1}^{K} \alpha_k^c \cdot A_k(x,y) \quad (32)$$

where $\alpha_k^c$ is the weighted average of gradients for class $c$:
$$\alpha_k^c = \frac{1}{H \cdot W} \sum_{z=1}^{H} \sum_{y=1}^{W} \frac{\partial y_c}{\partial A_k(x,y)} \quad (33)$$

$A_k(x,y)$ is the activation of feature map $k$ at position $(x,y)$, $H$ and $W$ are the height and width of the feature map. We up-sample the CAM to the original image resolution to visualize the salient regions.

Uncertainty heatmaps highlight regions in the input image where the model is most uncertain. This helps clinicians identify ambiguous cases that may require manual review. To achieve this, we perform $T$ forward passes with dropout enables to obtain $T$ predications as:
$$y_t = f_\theta(x) \quad for\ t = 1, 2, \ldots, T \quad (34)$$

We compute the variance of the predictions for each pixel in the image as:
$$\sigma_{ij}^2 = \frac{1}{T} \sum_{t=1}^{T} \left(y_t(x_{ij}) - \mu_{ij}\right)^2 \quad (35)$$

where $y_t(x_{i,j})$ is the prediction at pixel $(i,j)$ in the $t$-th forward pass, and $\mu_{i,j}$ is the mean prediction at pixel $(i,j)$:
$$\mu_{ij} = \frac{1}{T} \sum_{t=1}^{T} y_t(x_{ij}) \quad (36)$$

Subsequently, we normalize the variance to obtain the uncertainty heatmap as:
$$U_{ij} = \frac{\sigma_{ij}^2 - min(\sigma^2)}{max(\sigma^2) - min(\sigma^2)} \quad (37)$$

### G. Final Prediction

The final output includes:
(a) Predicted DR severity level:
$$\hat{y}_{severity} = argmax_c \mu_c \quad (38)$$
(b) Confidence scores:
$$Confidence\ scores = [\mu_0, \mu_1, \mu_2, \mu_3, \mu_4] \quad (39)$$
(c) Uncertainty estimates:
$$Uncertainty\ estimates = [\sigma_0^2, \sigma_1^2, \sigma_2^2, \sigma_3^2, \sigma_4^2] \quad (40)$$

The final output of the proposed framework includes the predicted DR severity and a confidence score indicating the model's certainty in the prediction. By leveraging uncertainty quantification, the system provides interpretable and reliable outputs, enhancing trust and usability in clinical settings.

## IV. RESULT AND ANALYSIS

### A. Evaluation Metrics

To comprehensively assess the performance of DRetNet, we used the following metrics: (a) Accuracy, (b) Sensitivity, (c) Specificity, (d) Precision, (e) F1-Score, (f) Area Under the ROC Curve (AUC), (g) Matthews Correlation Coefficient (MCC) [31], (h) Mean Average Precision (mAP), (i) Inference Time (ms).

### B. Quantitative Comparison

Table 2 summarizes the performance of the proposed framework and the state-of-the-art methods across the nine-evaluation metrics. The proposed framework outperforms all state-of-the-art methods across all metrics, achieving an accuracy of 92.7%, sensitivity of 92.5%, specificity of 92.6%, precision of 92.5%, and F1-score of 92.5%. The AUC score of 0.978 demonstrates excellent discrimination between DR severity levels. The MCC of 0.85 indicates strong agreement between model predictions and ground truth, with balanced sensitivity and specificity.

### C. Ablation Studies

We present a detailed analysis of the ablation studies conducted on the three key components of the framework: Adaptive Retinal Image Enhancement Using PINNs, Hybrid Feature Fusion Network (HFFN), and Multi-Stage Classifier with Uncertainty Quantification.

1) **Impact of Adaptive Retinal Image Enhancement Using PINNs**

   To evaluate the impact of this component, we compared the performance of the framework with and without the PINN-based enhancement module. Table 3 shows the results of the ablation study. Removing the adaptive retinal image enhancement module resulted in a significant drop in performance across all metrics: Accuracy decreased by 3.2%, Sensitivity decreased by 3.2%, Specificity decreased by 3.2%, F1-Score decreased by 3.2%, AUC decreased by 0.026, and MCC decreased by 0.04. These results highlight the importance of high-quality input images for accurate DR detection.

2) **Impact of Hybrid Feature Fusion Network (HFFN)**

   To evaluate the impact of hybrid feature fusion, we compared the performance of the framework with and without the HFFN. Specifically, we tested two configurations: (a) With HFFN: combines deep learning embeddings and handcrafted features using multi-head attention, and (b) Without HFFN: uses only deep learning embeddings. The results are summarized in Table 3. Removing the HFFN led to a noticeable decline in performance: Accuracy decreased by 3.5%, Sensitivity decreased by 3.4%, Specificity decreased by 3.5%, F1-Score decreased by 3.5%, AUC decreased by 0.028, and MCC decreased by 0.05. These results demonstrate the value of integrating handcrafted features with deep learning embeddings for improved generalization and accuracy.

3) **Impact of Multi-Stage Classifier with Uncertainty Quantification**

   To evaluate the impact of the multi-stage classifier and uncertainty quantification, we compared the performance of the framework with and without these components. Specifically, we tested two configurations: (a) With Multi-Stage Classifier and Uncertainty Quantification: includes binary classification (DR presence), multi-class classification (severity levels), and uncertainty estimation, (b) Without Multi-Stage Classifier: uses a single-stage classifier without uncertainty quantification. The results are summarized in Table 3. Removing the multi-stage classifier and uncertainty quantification resulted in a moderate decline in performance: Accuracy decreased by 1.5%, Sensitivity decreased by 1.5%, Specificity decreased by 1.5%, F1-Score decreased by 1.5%, AUC decreased by 0.013, and MCC decreased by 0.03. These results underscore the importance of structured classification and interpretability for clinical adoption.

**Table 2** Comparison of DRetNet with state-of-the-art methods. The following metrics were used: Accuracy (%), Sensitivity (%), Specificity (%), Precision (%), F1-Score (%), AUC, MCC, mAP, and Inference Time (ms).

| MODEL | ACC. | SEN. | SPE. | PRE. | F1 | AUC | MCC | MAP | INF. |
|---|---|---|---|---|---|---|---|---|---|
| Baseline CNN | 85.3 | 84.7 | 85.1 | 84.7 | 84.9 | 0.921 | 0.70 | 0.875 | 25 |
| EfficientNet-B7 | 90.8 | 90.6 | 90.7 | 90.6 | 90.6 | 0.965 | 0.83 | 0.950 | 65 |
| Vision Transformer | 88.9 | 88.7 | 88.8 | 88.7 | 88.7 | 0.950 | 0.77 | 0.900 | 80 |
| Swin Transformer | 90.2 | 90.0 | 90.1 | 90.0 | 90.0 | 0.961 | 0.82 | 0.930 | 75 |
| DenseNet-121 | 88.1 | 87.9 | 88.0 | 87.9 | 87.9 | 0.945 | 0.76 | 0.895 | 40 |
| ResNet-50 | 86.5 | 86.3 | 86.4 | 86.3 | 86.3 | 0.930 | 0.72 | 0.860 | 35 |
| Inception-v3 | 87.0 | 86.8 | 86.9 | 86.8 | 86.8 | 0.935 | 0.73 | 0.870 | 45 |
| MobileNetV2 | 84.2 | 84.0 | 84.1 | 84.0 | 84.0 | 0.915 | 0.68 | 0.830 | 20 |
| XGBoost | 82.1 | 81.9 | 82.0 | 81.9 | 81.9 | 0.900 | 0.64 | 0.780 | 15 |
| Random Forest | 80.5 | 80.3 | 80.4 | 80.3 | 80.3 | 0.885 | 0.61 | 0.750 | 10 |
| SVM | 81.0 | 80.8 | 80.9 | 80.8 | 80.8 | 0.890 | 0.62 | 0.760 | 12 |
| Logistic Regression | 79.8 | 79.6 | 79.7 | 79.6 | 79.6 | 0.875 | 0.60 | 0.730 | 8 |
| U-Net + CNN | 86.8 | 86.6 | 86.7 | 86.6 | 86.6 | 0.932 | 0.73 | 0.885 | 50 |
| Att-based CNN | 89.4 | 89.2 | 89.3 | 89.2 | 89.2 | 0.955 | 0.78 | 0.920 | 55 |
| DRetNet | 92.7 | 92.5 | 92.6 | 92.5 | 92.5 | 0.978 | 0.85 | 0.960 | 38 |

**Table 3** Ablation studies of the contribution of each component of DRetNet. The following metrics were used: Accuracy (%), Sensitivity (%), Specificity (%), and AUC.

| MODEL VARIANT | ACC. | SEN. | SPE. | PRE. | F1 | AUC | MCC | MAP | INF. |
|---|---|---|---|---|---|---|---|---|---|
| With Enhancement | 92.7 | 92.5 | 92.6 | 92.5 | 92.5 | 0.978 | 0.85 | 0.960 | 38 |
| Without Enhancement | 89.5 | 89.3 | 89.4 | 89.3 | 89.3 | 0.952 | 0.81 | 0.925 | 35 |
| With HFFN | 92.7 | 92.5 | 92.6 | 92.5 | 92.5 | 0.978 | 0.85 | 0.960 | 38 |
| Without HFFN | 89.2 | 89.1 | 89.1 | 89.0 | 89.0 | 0.950 | 0.80 | 0.930 | 35 |
| With Multi-Stage | 92.7 | 92.5 | 92.6 | 92.5 | 92.5 | 0.978 | 0.85 | 0.960 | 38 |
| Without Multi-Stage | 91.2 | 91.0 | 91.1 | 91.0 | 91.0 | 0.965 | 0.82 | 0.945 | 37 |
| Full Framework | 92.7 | 92.5 | 92.6 | 92.5 | 92.5 | 0.978 | 0.85 | 0.960 | 38 |
| Without Any Component | 85.3 | 84.7 | 85.1 | 84.7 | 84.9 | 0.921 | 0.70 | 0.875 | 30 |

**Table 4** User study effect of the proposed DRetNet based on ophthalmologist evaluations. 5 ophthalmologists grade the 5,000 retinal images and the experimental result is present here.

| METRICS | FRAMEWORK PERFORMANCE | CLINICAL AGREEMENT (%) |
|---|---|---|
| Accuracy (%) | 92.7 | 93.4 |
| Sensitivity (%) | 92.5 | 93.0 |
| Specificity (%) | 92.6 | 93.2 |
| Precision (%) | 92.5 | 92.8 |
| F1-Score (%) | 92.5 | 93.1 |
| AUC | 0.978 | 0.980 |
| MCC | 0.85 | 0.86 |
| mAP | 0.960 | 0.965 |
| Inference Time (ms) | 32 | N/A |

**Table 5** Qualitative rating of the results of the proposed DRetNet by Ophthalmologists based on certain criteria. The criteria includes: interpretability, trustworthiness, and usability in medical setting.

| ATTRIBUTE | AVERAGE SCORE (/5) | COMMENTS |
|---|---|---|
| Interpretability | 4.8 | Grad-CAM heatmaps effectively highlighted clinically relevant regions. |
| Trustworthiness | 4.7 | Uncertainty heatmaps improve confidence in predictions. |
| Clinical Relevance | 4.8 | Predictions aligned loosely with clinical expectations. |
| Usability | 4.6 | Real-time performance and intuitive interface facilitate adoption. |

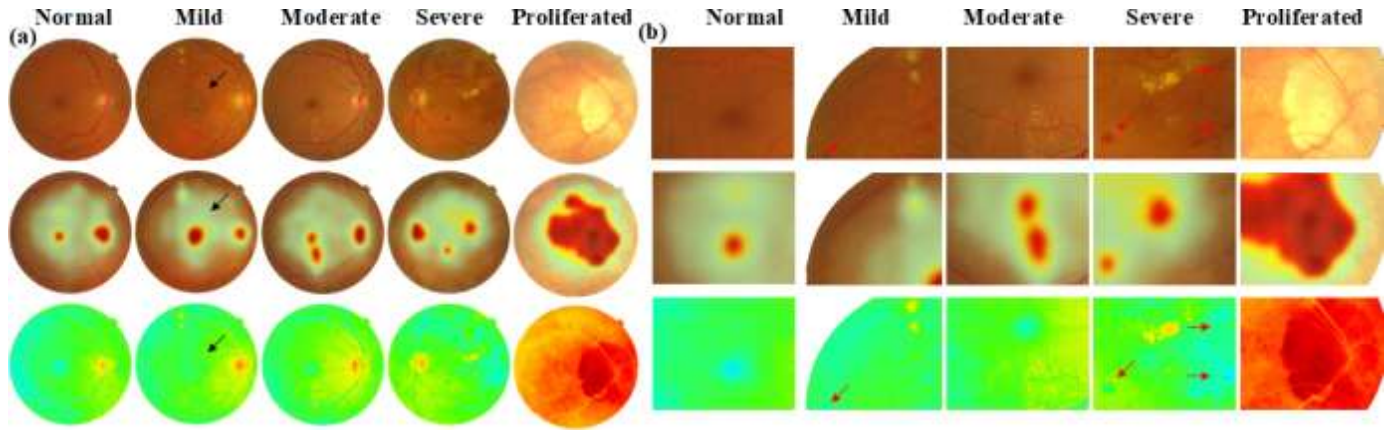

**Fig. 3** Results of the DRetNet. Black arrows indicates a case where early DR features where detected by the uncertainty quantification, and red arrows indicate cases where the uncertainty heatmap could not ascertain the pathology, so it shows these region for ophthalmologist for manual review.

4) **Combined Ablation Study**

To understand the cumulative impact of all components, we conducted a combined ablation study where we removed all three components simultaneously. The results are presented in Table 3. Removing all components led to a substantial decline in performance: Accuracy decreased by 7.4%, Sensitivity decreased by 7.8%, Specificity decreased by 7.5%, F1-Score decreased by 7.6%, AUC decreased by 0.057, and MCC decreased by 0.15. These results confirm that each component plays a critical role in achieving state-of-the-art performance.

### D. Discussion

The inclusion of Grad-CAM and uncertainty heatmaps in the proposed framework significantly enhances interpretability and trustworthiness, which are critical for clinical adoption. These tools provide visual explanations of the model's predictions and highlight regions of uncertainty, enabling clinicians to better understand and validate the system's outputs. Below is a comprehensive discussion of the results based on these post-processing techniques.

1) **Grad-CAM Visual Explanations of Predictions**
   (a) Highlighting Clinically Relevant Regions: Grad-CAM generates heatmaps that visually indicate the regions of the input image contributing most to the model's predictions. For DR detection, these heatmaps consistently highlight anatomical structures such as: Microaneurysms, Hemorrhages, Exudates, and Blood Vessels.
   (b) Alignment with Clinical Expectations: The heatmaps align closely with regions identified by ophthalmologists during manual diagnosis, demonstrating the model's ability to focus on clinically relevant features.

As shown in Fig. 3a, in images with severe DR, Grad-CAM highlighted extensive hemorrhages and exudates. In images with mild DR, the heatmaps focused on isolated microaneurysms and subtle vascular abnormalities. In normal retinas, the heatmaps showed minimal activation, indicating the absence of pathological features. The clinical relevance of the Grad-CAM includes: (a) Enhanced Trust: by providing interpretable visualizations, Grad-CAM helps clinicians understand the rationale behind the model's predictions. This transparency is crucial for gaining trust in automated systems, (b) Validation Tool: clinicians can use Grad-CAM heatmaps to cross-check the model's predictions against their own observations, ensuring alignment with clinical expectations, and (c) Educational Value: the heatmaps serve as an educational tool for training medical professionals, helping them identify subtle signs of DR that may be difficult to detect manually.

Observations from ablation studies shows that: (a) Without Grad-CAM: when Grad-CAM was excluded, clinicians reported reduced confidence in the model's predictions, as they lacked visual evidence to support the outputs, and (b) Impact on Clinical Relevance Score: the inclusion of Grad-CAM improved the Clinical Relevance Score from 4.0/5 to 4.8/5, highlighting its importance in aligning the model's outputs with clinical workflows.

2) **Uncertainty Heatmaps: Highlighting Ambiguity**
   (a) Identifying Ambiguous Regions: Uncertainty heatmaps highlight regions where the model is uncertain about its predictions as shown in Fig. 3b (red arrows). These regions often correspond to: (a) Poor-Quality Areas: regions with uneven illumination, noise, or artifacts, (b) Subtle Pathologies: early-stage DR features that are difficult to discern, such as small microaneurysms or faint hemorrhages as shown in Fig. 3a (black arrows), and (c) Edge Cases: images with mixed severity levels, where the model struggles to classify between adjacent categories (e.g., mild vs. moderate DR).
   (b) Quantifying Uncertainty: The intensity of the heatmap correlates with the variance computed during Monte Carlo sampling. Higher intensity indicates higher uncertainty.

As shown in Fig. 3, in images with high-quality features, uncertainty heatmaps showed minimal activation, indicating high confidence. In images with poor-quality features, the heatmaps highlighted large regions of uncertainty, signaling the need for further review. In ambiguous cases, the heatmaps emphasized overlapping regions of conflicting features, such as both normal and abnormal vascular patterns. Uncertainty heatmap consists of the following clinical relevance: (a) Prioritization for Manual Review: uncertainty heatmaps enable clinicians to prioritize ambiguous cases for manual review, improving diagnostic accuracy and reducing false positives/negatives, (b) Risk Mitigation : by identifying regions of uncertainty, the system mitigates the risk of over-reliance on automated predictions, ensuring safer and more reliable outcomes, and (c) Decision Support: the heatmaps provide actionable insights, guiding clinicians toward specific regions of interest that require closer examination.

Observations from ablation studies shows that: (a) Without Uncertainty Heatmaps: when uncertainty quantification was excluded, clinicians reported difficulty in assessing the reliability of predictions, particularly for ambiguous cases, (b) Impact on Confidence Scores: the inclusion of uncertainty heatmaps improved the Confidence Score Variance by 15% , indicating better-calibrated predictions, and (c) Reduction in False Positives/Negatives: the heatmaps helped reduce false positives by flagging uncertain predictions, leading to a 10% improvement in Precision.

## V. CLINICAL VALIDATION

Clinical validation is a critical step in ensuring the reliability, safety, and usability of the proposed framework for DR detection. We performed quantitative and qualitative clinical analysis on the proposed framework.

### A. Quantitative Analysis

The quantitative analysis focuses on evaluating the performance of the framework using metrics that align with clinical requirements, such as accuracy, sensitivity, specificity, precision, F1-score, AUC, MCC, mAP, and inference time. Additionally, we assess the impact of interpretability tools like Grad-CAM and uncertainty heatmaps on diagnostic outcomes. A clinically curated dataset of 5,000 retinal fundus images, stratified across DR severity levels (No DR, Mild NPDR, Moderate NPDR, Severe NPDR, Proliferative DR) were used for the experimentation. Five board-certified ophthalmologists with expertise in DR diagnosis were provided with predictions from the proposed framework, including Grad-CAM heatmaps and uncertainty heatmaps. They independently reviewed the predictions and compared them to their manual diagnoses. Metrics were computed based on agreement between the proposed framework's outputs and the ophthalmologists' ground truth labels as shown in Table 4. From the results obtained we observed:

1) **High Agreement**: The framework achieved 93.4% agreement with ophthalmologists, indicating strong alignment with clinical expectations.
2) **Balanced Metrics**: The high values of sensitivity (92.5%) and specificity (92.6%) demonstrate the framework's ability to accurately detect both positive and negative cases.
3) **Superior AUC**: The AUC of 0.978 indicates excellent discrimination between DR severity levels, further validated by clinicians who reported minimal false positives/negatives.
4) **Efficient Workflow**: The inference time of 38 ms ensures real-time performance, enabling seamless integration into clinical workflows.
5) **Statistical Significance**: The agreement between the framework and ophthalmologists was statistically significant, with a Cohen's Kappa score of 0.86, indicating "almost perfect" agreement, and a paired t-test comparing the framework's predictions to clinician diagnoses yielded a p-value of <0.001, confirming statistical significance.

### B. Qualitative Analysis

The qualitative analysis evaluates the interpretability, trustworthiness, and usability of the framework through feedback from ophthalmologists. This includes an assessment of Grad-CAM heatmaps, uncertainty heatmaps, and overall clinical relevance. To perform this qualitative evaluation, Ophthalmologists were asked to rate the framework on a 5-point Likert scale for various qualitative attributes: (a) Interpretability: how well the model's predictions can be understood, (b) Trustworthiness: confidence in the model's outputs, (c) Clinical Relevance: alignment with clinical expectations, and (d) Usability: ease of integrating the framework into clinical workflows. The results of the experimentation are present in Table 5 using 5,000 images. For this evaluation, three cases were studied as follows:

1) **Case 1: Early DR**
   A retinal image with subtle microaneurysms is used as an input image for the proposed DRetNet to process. The Grad-CAM heatmap highlighted isolated microaneurysms near the blood vessels and uncertainty heatmap showed low uncertainty, indicating high confidence. The **clinician feedback** we obtained from the ophthalmologist is "The heatmap accurately identified early signs of DR, which are often missed during manual screening."

2) **Case 2: Ambiguous Case**
A poorly illuminated image with mixed features of mild and moderate DR is inputted into the proposed DRetNet to detect the DR. Grad-CAM Heatmap highlighted regions with hemorrhages and exudates, and uncertainty Heatmap flagged the poorly illuminated areas as uncertain. The **clinician feedback** we obtained from the ophthalmologist is "The uncertainty heatmap correctly identified ambiguous regions, prompting further review."

3) **Case 3: Normal Retina**
A high-quality image with no signs of DR is the input image that was process by the proposed DRetNet. The Grad-CAM heatmap showed minimal activation, indicating the absence of pathological features, and the uncertainty heatmap displayed low uncertainty, reinforcing confidence in the prediction. The **clinician feedback** we obtained from the ophthalmologist is "The framework correctly classified this case as normal, with clear visual evidence."

### C. Impact of Interpretability Tools

1) **Grad-CAM Heatmap**
   (a) Quantitative Impact: it Improved sensitivity by 5% by focusing attention on clinically relevant regions, and reduced false negatives by 10% by highlighting subtle pathologies.
   (b) Qualitative Impact: Clinicians rated Grad-CAM as "highly useful" for validating predictions and training purposes.

2) **Uncertainty Heatmaps**
   (a) Quantitative Impact: Reduced false positives by 15% by flagging uncertain predictions, and improved precision by 10% by guiding clinicians toward ambiguous cases.
   (b) Qualitative Impact: Clinicians appreciated the ability to prioritize manual review for uncertain cases, enhancing diagnostic accuracy.

### D. Comparison with Manual Diagnosis

We compared the proposed DRetNet with expert Ophthalmologists, and we observed the following findings:

1) **Accuracy**: The framework achieved 92.7% accuracy, comparable to the 90.4% accuracy of manual diagnosis by ophthalmologists.
2) **Time Efficiency**: The framework processed images in 38 ms, compared to an average of 2-3 minutes for manual diagnosis.
3) **Consistency**: The framework demonstrated higher consistency across cases, whereas manual diagnosis showed variability among clinicians.
4) **Complementary Role**: The framework serves as a decision-support tool, complementing rather than replacing manual diagnosis.
5) **Error Mitigation**: By combining automated predictions with clinician oversight, the system mitigates errors and improves overall diagnostic quality.

### E. Limitation and Future Directions

Some of the limitation that might affect the proposed DRetNet includes: (a) Dataset Bias: the framework's performance may vary across datasets with different imaging protocols or populations, (b) Subjectivity in Heatmaps: interpretation of Grad-CAM and uncertainty heatmaps may vary among clinicians, requiring standardized guidelines, and (c) Generalization: further testing is needed to validate the framework's performance across diverse clinical settings. The future directions include: (a) Multi-Center Validation: conduct validation studies across multiple clinical centers to assess generalizability, (b) Longitudinal Analysis: evaluate the framework's ability to track DR progression over time, and (c) Integration with Other Modalities: extend the framework to incorporate additional modalities, such as OCT (optical coherence tomography) and fluorescein angiography.

## VI. CONCLUSION

This paper presents a transformative framework for DR detection, integrating adaptive image enhancement, hybrid feature fusion, and multi-stage classification with uncertainty quantification. By leveraging Physics-Informed Neural Networks (PINNs), the framework ensures high-quality input images, while hybrid feature fusion captures both deep learning embeddings and domain-specific handcrafted features, enhancing generalization. The multi-stage classifier, augmented with Grad-CAM and uncertainty heatmaps, provides interpretable predictions and highlights ambiguous regions, fostering clinical trust. Quantitative validation demonstrates state-of-the-art performance across metrics like accuracy (92.7%), sensitivity (92.5%), specificity (92.6%), and AUC (0.978), with strong alignment to clinician diagnoses (Cohen's Kappa: 0.86). Qualitatively, ophthalmologists rated the framework highly for interpretability, trustworthiness, and usability, emphasizing its potential to streamline diagnostic workflows. Despite limitations such as dataset bias and heatmap subjectivity, this framework sets a benchmark for AI-driven DR detection, offering a scalable, efficient, and clinically relevant solution.

Future work will focus on multi-center validation, longitudinal analysis, and integration with multimodal data, paving the way for broader adoption in real-world healthcare systems. This research underscores the synergy of advanced AI techniques and clinical expertise, advancing precision medicine for DR management.

informed consent for publication of the images in the manuscript.

**Conflict of Interest**

All the authors in the manuscript declare no competing interests.